\documentclass[useAMS,usenatbib,usegraphicx]{mn2e}
\usepackage{times}

\newif\ifAMStwofonts
\AMStwofontstrue

\newcommand{\simlt}{\lower.5ex\hbox{$\; \buildrel < \over \sim \;$}}
\newcommand{\simgt}{\lower.5ex\hbox{$\; \buildrel > \over \sim \;$}}
\newcommand{\be}{\begin{equation}}
\newcommand{\ba}{\begin{eqnarray}}
\newcommand{\ee}{\end{equation}}
\newcommand{\ea}{\end{eqnarray}}

\title[Dissecting the size evolution of elliptical galaxies since z$\sim$1]
{Dissecting the size evolution of elliptical galaxies since z$\sim$1:
  \\
Puffing up vs minor merging scenarios}
\author[Trujillo, Ferreras \& de la Rosa]
{Ignacio Trujillo$^{1,2}$, Ignacio Ferreras$^3$, Ignacio G. de la Rosa$^{1,2,4}$\\
$^1$ Instituto de Astrof\'\i sica de Canarias, C/ V\'\i a L\'a ctea
s/n, La Laguna, E-38200 La Laguna, Tenerife, Spain\\ 
$^2$ Departamento de Astrof\'\i sica,
Universidad de La Laguna, E-38205 La Laguna, Tenerife, Spain \\
$^3$ Mullard Space Science Laboratory, University College London, 
Holmbury St Mary, Dorking, Surrey RH5 6NT\\
$^4$ Department of Physics and Astronomy, University College London, Gower
Street, London, WC1E 6BT}

\begin{document}
\date{Revised version, MN11-0345, April 18, 2011}
\pagerange{\pageref{firstpage}--\pageref{lastpage}} \pubyear{2011}
\maketitle
\label{firstpage}

\begin{abstract}
  At fixed stellar mass, the size of low redshift early-type
    galaxies is found to be a factor of two larger than their
    counterparts at z$\sim$1, a result with important implications to
    galaxy formation models.  In this paper, we have explored the
  buildup of the local mass-size relation of elliptical galaxies using
  two visually classified samples. At low redshift we compiled a
  subsample of 2,656 elliptical galaxies from the Sloan Digital Sky
  Survey (SDSS), whereas at higher redshift (up to z$\sim$1) we
  extracted a sample of 228 objects from the {\sl HST/ACS} images of
  the Great Observatories Origins Deep Survey (GOODS). All the
  galaxies in our study have spectroscopic data, allowing us to
  determine the age and mass of the stellar component. Contrary to
  previous claims in the literature, using the fossil record
  information contained in the stellar populations of our local
  sample, we do not find any evidence for an age segregation at a
  given stellar mass depending on the size of the galaxies. At a fixed
  dynamical mass there is only a $\simlt$9\% size difference in the
  two extreme age quartiles of our sample. Consequently, the local
  evidence does not support a scenario whereby the present-day
  mass-size relation has been progressively established via a
  bottom-up sequence, where older galaxies occupy the lower part this
  relation, remaining in place since their formation.  We do not find
  any age segregation difference in our high-z sample
  either. Therefore, we find a trend in size that is insensitive to
  the age of the stellar populations, at least since z$\sim$1. This
  result supports the idea that the stellar mass-size relation is
  formed at z$\sim$1, with all galaxies populating a region which
  roughly corresponds to 1/2 of the present size distribution.  We
  have explored two possible scenarios for size growth: puffing up or
  minor merging. The fact that the evolution in size is independent of
  stellar age, together with the absence of an increase in the scatter
  of the relationship with redshift does not support the puffing up
  mechanism. The observational evidence, however, can not reject at
  this stage the minor merging hypothesis. We have made an estimation
  of the number of minor merger events necessary to bring the high-z
  galaxies into the local relation compatible with the observed size
  evolution. Since z=0.8, if the merger mass ratio is 1:3 we estimate
  $\sim$3$\pm$1 minor mergers and if the ratio is 1:10 we obtain
  $\sim$8$\pm$2 events.
\end{abstract}



\section{Introduction}

Present-day galaxies show a clear correlation between mass and size,
with the most massive galaxies having the larger sizes. This mass-size
relationship has been known both for elliptical and spiral galaxies
for many years. With the advent of large surveys, like the Sloan
Digital Sky Survey \citep[SDSS, ][]{sdss} it has been possible to
quantify this correlation with high accuracy \citep[see
e.g. ][]{she03}.  However, the mechanisms by which this relationship
is built remain uncertain. For instance, we do not have conclusive
answers to questions like: ``were the galaxies born in-situ at the
positions where we find them in the local mass-size relation or were
they born in another part on this diagram, drifting to their present
location?''. If so, ``how much have they grown and what are the
mechanisms responsible for this displacement?''. Answering these
questions is directly connected to our understanding of how the
assembly of the galaxies has proceeded through cosmic time. In this
paper we will particularly focus on spheroidal galaxies as it has
been shown in the last few years that their stellar mass-size relation
has dramatically changed with redshift.

Several papers have explored the evolution of the stellar mass-size
relation of spheroid--like galaxies
\citep[e.g. ][]{truj04,mci05,truj06a,truj07,bui08,ig09b,sar11}. In
general, they all agree with a significant evolution of this relation
with redshift. Their results can be summarized as follows: at fixed
stellar mass, spheroid-like galaxies were significantly more compact
at higher redshift
\citep[e.g. ][]{dad05,truj06b,lon07,zirm07,vdwel08,vdk08,cim08,dam09,car10},
with an increase on the effective radii by a factor of $\sim$2(4) from
z$\sim$1(2) \citep[e.g. ][]{truj07}. However, these observational
results say little about the amount of size evolution of individual
galaxies on the mass--size plane. Nevertheless, at least a few basic
statements can be established regarding the growth of individual
galaxies based on the current observational evidence. First, at high-z
there are no big spheroidal objects, implying that the present-day
large elliptical galaxies have either formed recently (in-situ) with
large sizes or they are the product of the evolution of previous
compact galaxies that populated the high-z stellar mass-size
plane. Second, the near absence of compact massive galaxies in the
nearby Universe \citep{truj09,val10,tay10}, which were very common in
the early Universe, indicates that individual objects (at least the
very old and compact ones) have evolved significantly in size.

Some recent works have conducted a detailed analysis of the buildup of
the local spheroid mass--size relationship \citep{vdwel09,val10}.
These works propose that the formation of this relation is a result of
two steps: a) the continuous emergence of galaxies as early-type
systems with larger sizes, as cosmic time increases, due to the
decreasing availability of gas during their formation phase
\citep{ks06}, and b) their subsequent growth through either gas
expulsion in the so-called puffing up scenario
\citep{fan08,fan10,dam09} or by minor merging activity
\citep{naab09,hop09}. If the above scenario is correct, i.e. that the
new assembled galaxies are born with larger sizes as redshift
decreases, we should observe that the number density of spheroid-like
massive galaxies at fixed stellar mass should decrease with increasing
redshift. Furthermore, a gradual change of the age of the galaxies at
fixed stellar mass should be expected, in the sense that larger
galaxies should be younger. However, there is no compelling evidence
of a significant drop in the number density of elliptical galaxies up
to z$\sim$1 \citep[see e.g.][]{ig09b}, weakening this formation
scenario. On the other hand, in \citet{vdwel08} and \citet{val10},
there is some hint that larger spheroid-like galaxies, at fixed
dynamical and stellar mass, are younger than their compact mass
equivalents.

In this paper we reexamine the buildup of the mass--size relationship
of spheroidal galaxies with two significant improvements in relation
to previous work. First, this paper addresses the issue of the
evolution of early-type galaxies on the size-mass plane by comparing a
nearby and a distant sample of galaxies, classified and analyzed in
the same way.  We will show in this paper that previous studies of the
local stellar mass--size relation of early-type galaxies are severely
contaminated by galaxies of other morphological types. For this
reason, the present study is the first one exploring objects that have
been classified only visually, and not by any other criteria, like
structural parameters or colours.  The second advantage of the present
work is that we have quality spectra for all our targets, allowing us
-- by exploring their spectral energy distributions (SEDs) -- to
obtain reliable star formation histories (SFHs). Spectroscopic data is
essential to robustly determine the properties (stellar mass and age)
of the underlying stellar populations in both local and high redshift
samples, allowing us to make a much more consistent assessment of the increase
in galaxy size on an individual galaxy basis. The information about
the ages allows us to explore whether the size evolution depends on the
properties of the stellar populations of these galaxies. This
information is key to distinguish between the two most likely mechanisms
of size growth proposed in the literature for elliptical galaxies:
the puffing up versus the minor merging scenario. We will amply
discuss in this paper the implications of our findings in
relation to these two models.

The paper is structured as follows. In Section 2 we describe the local
sample. The connection between stellar age and the mass--size
relationship is explored in Section 3. In Section 4 we present our
moderate redshift sample and in Section 5 we quantify the size
evolution of our galaxies. Section 6 is devoted to explore which
evolutionary scenario is more plausible according to these
observations, finally concluding on Section 7 with an overview of our
results. In this paper we adopt a standard $\Lambda$CDM cosmology, with
$\Omega_m$=0.3, $\Omega_\Lambda$=0.7 and H$_0$=70 km/s/Mpc.


\section{The local sample}

Our local sample is taken from the morphological catalogue of
\citet[][hereafter NA10]{na10} obtained via visual classification from
SDSS imaging. The NA10 catalogue comprises 14,034 galaxies from SDSS
Data Release 4 \citep[DR4,][]{sdssdr4} in the redshift range 0.01 $<$
z $<$ 0.1 with an extinction-corrected g-band magnitude brighter than
16. From this catalogue, we select the elliptical galaxies (c0, E0 and
E+) with a T-Type class -5, resulting in a final sample comprising
2,656 galaxies with available spectra.

\begin{figure*}
\includegraphics[width=16cm]{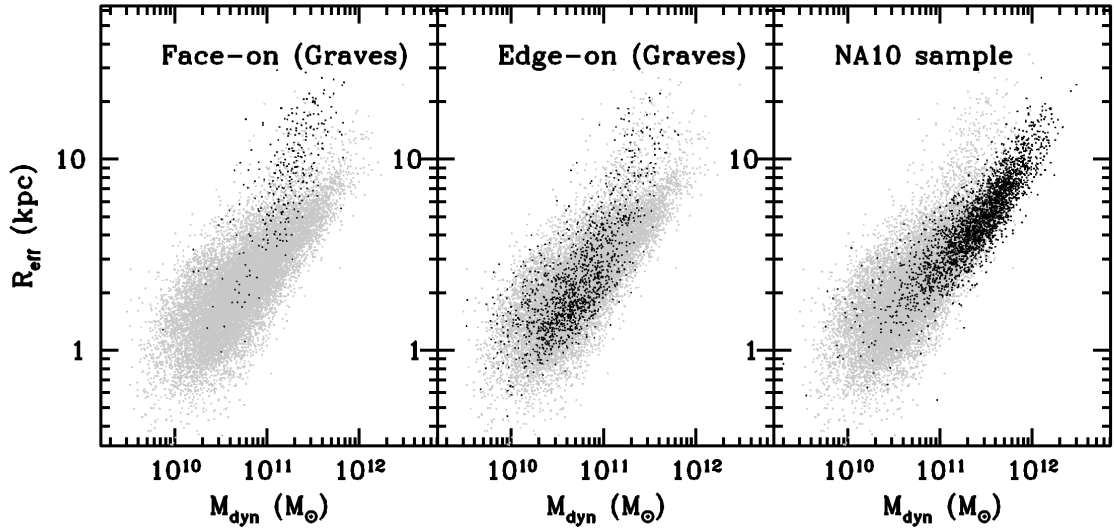}
 \caption{Dynamical mass--size relation of the sample used in this
  paper (NA10 sample; black dots in the right column) compared to the
  early--type sample selection of \citet[][grey data points]{grav08}.
  This figure illustrates the different position in the mass--size
  diagram that spiral galaxies (i.e. contaminants) have in this diagram
  (see text for details). Late-type galaxy contaminants (LTGs) are not 
  distributed homogeneously over the early-type galaxy footprint:
  face-on LTGs mainly live in the top part of the diagram (large radii),
  whereas edge-on LTGs populate the bottom-left corner (low sizes and
  masses).}
  \label{contamination}
\end{figure*}

Our choice for visually classified galaxies aims at minimizing the
impact of morphological contaminants, which frequently degrade
automated classification samples. To illustrate this issue, we compare
in Fig.~\ref{contamination} the NA10 sample used in this paper with
the \citet[]{grav08} early-type sample used in \citet[]{vdwel09} for a
similar analysis to the one presented here. In contrast with our
selection, which is purely based on morphology, the \citet[]{grav08}
galaxies are selected by the following criteria: a) located on the red
sequence, b) no emission lines in their spectra and c) concentration
parameter C$>$2.5. From the comparison between both samples shown in
Fig.~\ref{contamination}, it follows that a sample of early-type
galaxies based on the above criteria will be contaminated by bulges of
both face-on and edge-on spirals. Unfortunately, those contaminants
are not distributed evenly throughout the sample. Instead, they tend
to concentrate in certain regions of the parameter space, introducing
undesired systematic effects.  For example, we see that the largest
galaxies in the mass--size plane, at a fixed mass, are heavily
contaminated by spiral galaxies. Another important advantage of our
selection purely based on morphology is that it is neither biased
against galaxies with recent star formation activity, nor with a
passive evolution in their star formation history. This is relevant
for the purpose of this paper as shown below.  We take advantage of
the fact that virtually all of the galaxies in the sample of
\citet{grav08} have their morphology visually studied by the Galaxy
Zoo project \citep{lint11}. We use their {\it fraction of votes} for
ellipticals ($p_{el}$); spirals (both clock-wise and
anti-clock-wise, $p_{sp}$) and edge-on spirals ($p_{edge}$) to
identify the Late Type Galaxy (LTG) contaminants. We have carried out
a further visual check of these contaminants, finding a complete
agreement with the results of Galaxy Zoo.  Despite the small overall
contamination rate ($\sim$1.8\%), the face-on LTGs concentrate in the
region with large radii, while the edge-on LTGs ($\sim$8\%)
concentrate towards low values of R$_e$ and $M_{\rm dyn}$.

The spectroscopic data and photometric parameters of the NA10 sample
are retrieved from the SDSS archive. We have used spectra from
DR7\citep{dr7}, to benefit from the improved flux calibration
introduced in DR6 \citep[see][]{sdssdr6}. The SDSS spectroscopic data
cover a wavelength range from roughly 3,800 to 9,200 \AA\ at an
average spectral resolution of 3.25 \AA\ (FWHM). This instrumental
resolution is not constant but varies in a complex way with
wavelength, fiber and arrangement. All spectra are both de-redshifted
and corrected for Galactic foreground extinction, using the dust maps
of \citet{Sch98}.  Hereafter, all size estimates are quoted as the
circularized effective radius $R_e \equiv(b/a)^{1/2}\times R_{\rm
  deV}$, with parameters deVRad\_g and deVAB\_g taken from the
photometric SDSS-pipeline. In principle, velocity dispersion data
($\sigma$) are also available from the DR7 SDSS-pipeline, although
with a moderately high ratio of missing values, amounting to over 15\%
of our SDSS sample. Consequently, we have re-calculated the values of
velocity dispersion with the same spectral fitting method used in this
study (STARLIGHT, see \S3), taking as velocity dispersion the
smoothing parameter of the stellar population mixture that produces
the best fit to the observed spectrum. \citet{LB10} show that
there is good agreement between the STARLIGHT and SDSS-DR7 velocity
dispersion values, with only a small systematic trend at the low ($<$
90 km/s) and high ($>$ 280 km/s) ends of the $\sigma$ range. Very few
measurements (0.4 \%) are excluded, with $\sigma<$~40~km/s, because
they are considerably smaller than the resolution of the base SSP
models (58 km/s). We have used the \citet{Jo95} prescriptions to
correct the velocity dispersion to the same fraction of the effective
radius, R$_e$/8, instead of the fixed fiber diameter (3 arcsec).

\section{The local mass--size plane: distribution of galaxies
  according to their stellar age}

Both \citet[]{vdwel09} and \citet[]{val10} have argued that there is
an age gradient within the mass--size plane of early--type galaxies:
at fixed mass, galaxies with larger sizes are found to be younger. As
it was argued in \S1, this observation is expected if newly assembled
spheroidal galaxies feature larger sizes than those systems assembled
earlier. For this reason, we have revisited the age distribution of
our galaxies in the NA10 sample.

The age of the stellar populations of our galaxies is estimated as
follows. We use the spectral fitting code STARLIGHT \citep{CF05} to
find combinations of single stellar population (SSP) models that,
broadened with a given velocity dispersion, achieve the best match
with the observed galaxy spectrum. For the present study, we have used
the spectral energy distributions of the MILES SSP models \citep{Va10}
with a Kroupa Universal Initial Mass Function \citep{K01}. These
models are based on the MILES\footnote{\tt www.iac.es/proyecto/miles}
stellar library \citep{SB06}, which combines both a rather complete
coverage of the stellar atmospheric parameters and a relatively high
and nearly constant spectral resolution, 2.3\AA\ (FWHM), optimally
suited for the spectral resolution of the SDSS data. Our base for the
fitting using STARLIGHT consists of 138 solar-scaled SSP models with 6
different metallicities, ranging from Z=1/50 to 1.6$\times$Z$_{\odot}$
and 23 different ages, from 0.08 to 11.22 Gyr. Extinction due to
foreground dust is modeled with the CCM-law \citep{CCM89} and masks
are used to avoid emission lines or bad pixels. The M$_{\rm stars}$
parameter -- the fraction of the initial stellar mass which still
remains as stars at a later time -- is extracted from the model
predictions and used to calculate the stellar mass of our galaxies,
M$_s$. In the present work, we have characterized the stellar
population mixture of each galaxy by its mass-weighted average age,
calculated as:

\begin{eqnarray*}
\langle {\rm age}\rangle_M & = & \displaystyle\sum_{j=1}^{N_*} \mu_j t_j
\end{eqnarray*}

where $N_*$ is the number of SSP models in the base, $\mu_j$ is the
{\it mass fraction vector}, defined as the fractional contribution of
the SSP with age $t_j$ and metallicity $Z_j$, to the total flux,
converted into mass with the M/L$_j$ of each SSP. Stellar masses,
M$_s$, are also computed with the $\mu_j$ and M/L$_j$. Once each SFH
has been calculated, the corresponding lookback-times are added in
order to set all the histories to a common z=0 ground. This offset
ranges between 0.1 (z=0.01) and 1.3~Gyr (z=0.1).

\begin{figure}
\includegraphics[width=8.8cm]{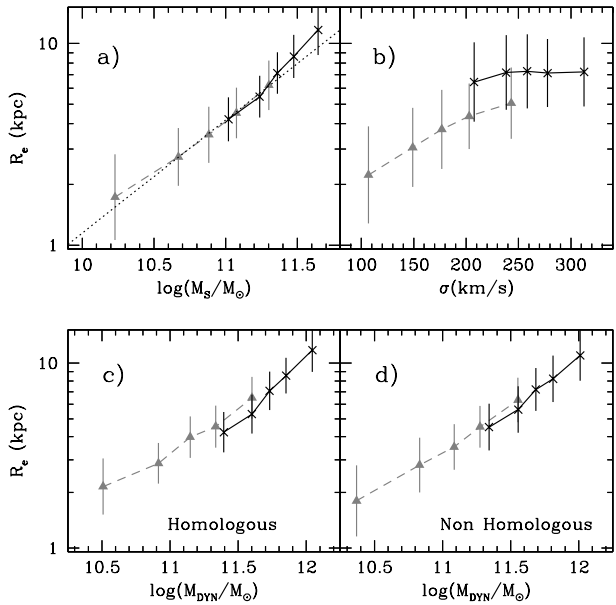}
 \caption{Correlation between effective radius and stellar mass (panel
  a), central velocity dispersion (panel b) and dynamical mass
  estimated assuming homology (panel c) and non-homology (panel d) for
  our local sample. The grey triangles (black crosses) represent the
  youngest (oldest) quartiles of the age (mass-weighted) distribution
  according to our modeling with STARLIGHT. The dotted line in 
  panel a) is the scaling relation of early-type galaxies according to
  \citet[]{she03}. }

  \label{fig:Sz0}
\end{figure}

\begin{figure}
\includegraphics[width=8.8cm]{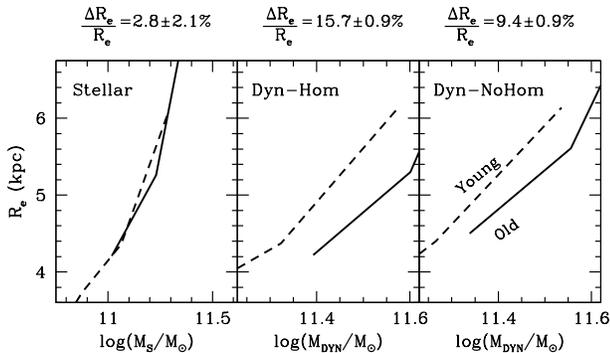}
 
\caption{Zoom into the mass-size distribution of our local sample. The
  regions shown are those where the two extreme age quartiles of the
  sample (young as a dashed line and old as a solid line) have
  galaxies over a similar range of stellar mass (left panel) and dynamical
  mass (middle and right panels). An estimate of the size difference in the overlapping
  regions are given in the top for each panel.}
  \label{fig:zoom}
\end{figure}

Figure~\ref{fig:Sz0} shows the size-mass correlation of our SDSS sample of early-type
galaxies, split according to the (mass-weighted) ages determined by STARLIGHT. We follow
the spirit of the modeling of \citet{vdwel08} and \citet{val10} whereby age-segregated
samples are expected to occupy different regions of the mass-size relation. In order to
maximize the difference according to age, we only show the upper and lower quartiles of
the age distribution, corresponding, respectively, to galaxies older than 11.7~Gyr (black
crosses), and younger than 10.2~Gyr (grey triangles). These symbols represent the median
within bins taken at a fixed number of galaxies per bin. The error bar gives the RMS
scatter within each bin. The figure shows that the predicted segregation in the mass-size
relation with respect to age is not significant when plotted against stellar mass (panel
a). This is in contradiction with the results shown in \citet[]{val10} (their Fig.~3). We
have quantified the separation between the young and the old galaxy families by zooming
in the region where the masses of the galaxies of our two extreme age quartiles overlap
(Fig.~\ref{fig:zoom}). We quantify the size difference as follows:
$\Delta$R$_e$/R$_e\equiv$2$\langle$R$_{e,{\rm young}}-$R$_{e,{\rm
old}}\rangle$/$\langle$R$_{e,{\rm young}}+$R$_{e,{\rm old}}\rangle$. As we expect from a
visual inspection of the figure, the size difference between the two families, at a fixed
stellar mass, is negligible (being compatible with zero change within the statistical
uncertainty). The reasons for this discrepancy could be several. On the one hand,
\citet{val10} segregates their galaxies using luminosity-weighted ages instead of
mass-weighted ages as used here. Another possibility is that their early-type selection
criteria based on the automatic code MORPHOT could include a larger number of spiral
galaxies as contaminants, in contrast with a visual classification. We have explored
whether using luminosity-weighted ages changes our results and find that this is not the
case. In fact, if we repeat the previous exercise using luminosity-weighted ages, we find
that the difference between the two extreme quartiles is 1.7$\pm$0.9\% (i.e. very similar
to the mass-weighted ages). Finally,  our stellar mass-size relation is compared with the
early-type relation of \citet{she03}.  The agreement is very good for objects with
stellar mass M$_s<3\times 10^{11}$M$\odot$. However, at the high mass end we note that
the sizes of our galaxies are slightly larger than those provided by \citet{she03}, a
result in agreement with \citet{guo09}, who find a similar underestimate of the Shen et
al. sizes of a similar sample of visually inspected early-type galaxies.

Our previous results show that at fixed stellar mass galaxies do not show any
significant difference in age.  However, an interesting change is found when dynamical
masses are considered instead of stellar ones (panels c and d). In this case, the age
segregation is apparent, with younger galaxies having slightly larger sizes. Under the assumption of dynamical homology (i.e. estimating the
dynamical masses as M$_{\rm dyn}=5\sigma^2R_e/G$ \citep{cap06}) the size difference among
the two extreme quartiles reaches a value of $\sim$16\% ($\sim$13\% in the case of
luminosity-weighted ages). However, elliptical galaxies are well known for not being an
homologous family. If we repeat the same analysis using this time the dynamical mass
accounting for the non-homology following the expression provided by \cite{bert02}:

\begin{equation}
M_{{\rm dyn},n}=K(n)\sigma^2R_{e,n}/G,
\label{eq:nonhom}
\end{equation}

\noindent
with

\begin{equation}
K(n)\simeq\frac{73.32}{10.465+(n-0.95)^2}+0.954,
\end{equation}

\noindent
and $n$ being the S\'ersic index of the elliptical galaxies in our sample
\citep[determined from ][]{blan05}, we find that the size difference,
$\Delta$R$_e$/R$_e$, decreases significantly to $\sim$9\% ($\sim$8\%
in the case of luminosity-weighted ages). Our findings about the size difference
 between the old and young galaxies at a fixed dynamical mass
is in qualitative agreement with the findings by \citet{vdwel09}.

Given that dynamical mass estimates depend on the velocity dispersion
quadratically, one would expect that the size difference of the
galaxies in the two age quartiles could be linked to a change on the
velocity dispersion between the young and the old population. Panel b)
of Fig.~\ref{fig:Sz0} confirms this point: at fixed
velocity dispersion, the older subsample is significantly larger
(35.3$\pm$0.5\%; 25.3$\pm$1.7\% in the case of luminosity-weighted
ages) than the younger galaxies.
Alternatively, one could interpret this result as follows: at fixed
effective radius, older galaxies have lower velocity dispersion than
their younger counterparts (although the region of overlap between
old and young galaxies at fixed size is arguably rather small). 

 Although we agree qualitatively with \citet{vdwel09} on a size difference between the
young and old galaxy families at a fixed dynamical mass, our results about the size
difference  at a fixed velocity dispersion is in contrast with them. These authors show
(in their Fig.~1) that at fixed velocity dispersion the age of the galaxies is
independent of their size. The reason for this discrepancy could be double: first, their
ages are luminosity-weighted, in contrast with our mass-weighted ages, and second, their
sample suffers from some contamination of spiral galaxies in key places of the mass-size
diagram. 

Irrespectively of the comparison with other works, our results indicate that the size
variation due to changes in the stellar population ages of the elliptical galaxies in the
local Universe is very small. Although the age trend goes in the direction (i.e. older
galaxies being more compact than young ones at fixed stellar mass) that one would expect
from a progressive bottom-up scenario for the buildup of the local mass-size relation
model, it is clear that the differences in size are very small to be able to reproduce
the large size variation with cosmic time found at high redshift. We will return to this
point more extensively in the following sections. We conclude that the stellar population
ages do not resemble the age of the full assembly of the elliptical galaxies and,
consequently, that {\sl after} the formation of the bulk of their stellar content,
elliptical galaxies have experienced a significant evolution in their size.

\subsection{Dynamical structure change of the galaxies with age}

The virial theorem predicts that, at fixed mass, the velocity
dispersion will change as the inverse of the root square of the galaxy
size. Consequently, one would expect, due to the strong size evolution
with redshift observed in the elliptical population, that the velocity
dispersion of the high redshift objects were significantly larger than
those found in local galaxies. However, observations are at odds with 
this scenario \citep{ct09}: the velocity dispersion of the elliptical
galaxies, at fixed stellar mass, only changes moderately with
redshift. \citet{hop09b} have explained this mild change in the
velocity dispersion suggesting that the contribution of the dark
matter halo to the gravitational potential of the galaxy changes
with cosmic time. According to that model, in the present Universe the
contribution of the dark matter halo on settling the velocity
dispersion of the galaxies will be higher than in the past.

\begin{figure}
\includegraphics[width=9cm]{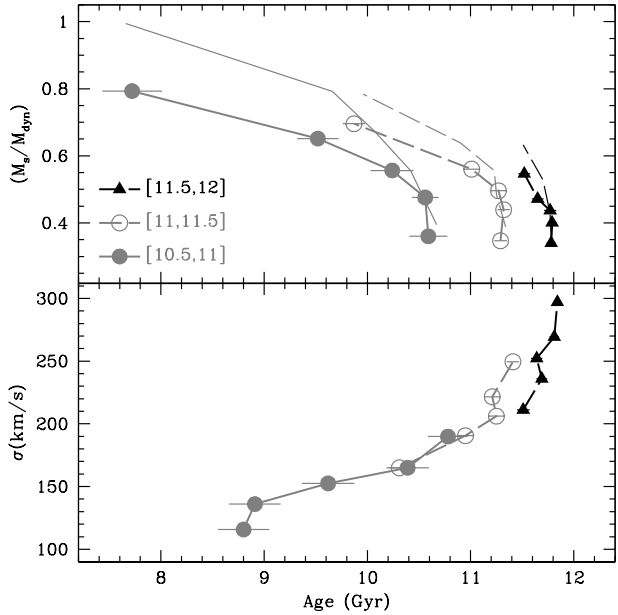}
\caption{The stellar mass-weighted age is shown with respect to baryon
  fraction (top) or velocity dispersion (bottom) for a range of
  stellar masses as labelled. The points and error bars give the
  median value and error in age bins chosen at fixed number of
  galaxies per bin. The thick lines indicate that the dynamical masses
  have been calculated assuming homology, whereas the
  mass estimates assuming non-homology are shown as thin lines.}
  \label{fig:tsigma}
\end{figure}

We can explore whether our local sample shows any hint of a dynamical
structure change as a function of the age as suggested by the
\citet{hop09b} idea. To do that we explore both the baryonic fraction
(top) and the velocity dispersion (bottom) of our local galaxies
against the age of their stellar populations in Fig.~\ref{fig:tsigma}.
The baryonic fraction is defined as the ratio between the stellar 
mass and the dynamical mass, and should roughly correspond to the net
baryon fraction within the effective radius for our early-type
galaxies. We bin the sample according to stellar mass as labelled.
 In order to illustrate the effect of non-homology, we show the
homologous estimates as thick lines and the non-homologous
models (using dynamical masses from Eq. \ref{eq:nonhom}) as thin lines.

The strong trend habitually found between the stellar populations and
velocity dispersion is evident, with the oldest galaxies being the
most massive ones \citep[see e.g.][]{ber05,grav08,rog10,nap10}, with a
larger dark matter content within the optical radius \citep[see
e.g.][]{fsw05,tor09,leier11}.  This trend of an increased dark matter
  content with galaxy mass is also consistent with the results
  pertaining to whole halos, as shown when comparing observed stellar
  mass functions with cosmological halo abundances \citep[see e.g.][]{mos10}.

We can now probe in more detail Fig.~\ref{fig:tsigma}: at fixed
stellar mass, the oldest galaxies feature higher velocity dispersions
and a lower baryon fraction. The higher velocity dispersion for the
older galaxies is in agreement with the findings of \citet{ct09} at
high redshift and supports the idea of \citet{hop09b}. However, the decrease
in the baryonic fraction as a function of age seems at odds with
the high-z findings. It is interesting to note that our estimation of
the dynamical mass is done with the sizes found in the local Universe,
so a direct relation with the baryonic fraction estimated at high-z is
not straightforward. From the analysis of the local relation, we find
that the age of the most massive local ellipticals is quite
homogeneous and also their dynamical structure change is limited to
M$_{s}$/M$_{\rm dyn}$$\sim$0.4$\pm$0.1. This suggests that the most
massive elliptical galaxies formed via an earlier, very homogeneous
formation process. This scenario is consistent with the observed lack
of evolution in the number density of massive early-type
galaxies \citep[see e.g.][]{fon06,ig09b,ban10}.

For our intermediate and lower stellar mass bin, elliptical galaxies
show a much more important trend between age and dynamical
structure. For instance, we see that for present
M$_{s}$$\sim$10$^{11}$M$_{\sun}$ ellipticals, the baryonic fraction
can change between 0.3 to 0.7 and the velocity dispersion between 150
to 250 km/s.  We note that our trend, at fixed stellar mass,
  towards a lower baryon fraction in older populations is at odds with
  \citet{shan09} and \citet{nap10}. They obtain the {\sl opposite}
  trend, namely more dark matter in the younger populations at fixed
  stellar mass \citep{nap10}, or corrected luminosity \citep{shan09}. However,
  our range of stellar masses and ages is much shorter, concentrated
  towards the high mass end. Furthermore, the age estimates of
  \citet{nap10} are based on broadband photometry alone, a method
  considered robust on the determinations of the stellar M/L but not
  on age estimates \citep[e.g.][]{fsw05}. \citet{shan09} use instead the
  spectroscopic ages from \citet{gal05} who use a combination of
  spectroscopic line strengths. In an independent study carried out
  with 40,000 ETGs from SDSS (de la Rosa et al. in preparation)
  several methods and SSP models are compared. The method of
  \citet{gal05} with \citet{BC03} models provide systematically
  younger ages than the {\it spectral-fitting} technique with the
  MILES population synthesis models used for the present
  study. Furthermore, by comparing the performance of the model-method
  combination with repeated observations of the same SDSS targets
  ($\sim$2300 repeated spectra), the {\it spectral-fitting} approach is shown to be
  considerably more robust than other age dating methods.
The difference between older and younger galaxies may
reflect different channels of galaxy formation within the same
  stellar mass bin.  As the size of the galaxies is proportional to
the dynamical mass and inversely proportional to the square of the
velocity dispersion, we obtain, as expected, a slight trend to smaller
galaxies as a function of the stellar population ages. This trend is
unable to explain the strong size evolution found at high redshift.

The detailed analysis of the local mass--size relation reveals that
the information contained is unable to fully explain which mechanisms
have followed the elliptical galaxies to reach their present
sizes. For this reason, it is necessary to conduct a direct comparison
of the properties of the local galaxies with those of equivalent
galaxies at high-z to extract such information. This is what we do in the
following sections.

\section{Moderate redshift sample}

In order to understand in more detail the size evolution of massive
galaxies and their relation to age, we include in our study a sample
of visually classified early-type galaxies at moderate redshift
(z$\simlt$1). The comparison with the local sample allows us to probe
the evolution of the mass-size relationship over the past
$\sim$8~Gyr. The deep images of the Great Observatories Origins Deep Survey
(GOODS) fields \citep{goods} taken by the Advanced Camera for Surveys
({\sl ACS}) on board the Hubble Space Telescope ({\sl HST}) provide
the optimal dataset for visual classification of galaxy morphologies
out to redshifts z$\simlt1$. We use the catalogue of early-type
galaxies from \citet{ig05} and \citet{ig09a} in the North and South
GOODS fields, comprising 910 visually classified early-type galaxies
brighter than F775W$=24$~mag (AB).  For a proper comparison with the
evolved local sample, we need a reliable estimate of stellar age.  The
broadband photometry of the GOODS sample is not good enough for our
purposes, and we consider a subsample with available spectral
data. The PEARS sample of early-type galaxies \citet{ig09c} comprises
228 galaxies from the GOODS catalogue, with available slitless
spectroscopy using grism G800L ({\sl HST/ACS}). The spectral
resolution depends on galaxy size, with an average value
R$\equiv\lambda/\Delta\lambda\sim$50 for our objects. This sample
covers a redshift range 0.4$<$z$<$1.3. The lower redshift was dictated
mainly by the requirement of having the 4000\AA\ break within the
sensitivity range of the grism data. In \citet{ig09c} stellar ages are
determined using a grid of composite models, including chemical
enrichment, from which best fit ages and metallicities are
obtained. However, in order to reduce the systematics, we only use
this modelling to generate the best fit spectra at similar resolution
to those from the local sample.  We note this method should introduce
a very small systematic given that the values of the reduced $\chi^2$
obtained for the PEARS sample are always of order one, and that the
method used in this paper to determine ages uses the full SED for
fitting, not individual absorption lines.

Ages and stellar masses are re-computed from these spectra, using the
same methodology as for the local sample \citep[i.e.
STARLIGHT][]{CF05}, with the only difference being the age range of
the model populations. For these galaxies we restrict the oldest SSPs
to the age of the Universe at the redshift of the galaxy.  This
approach is well justified as STARLIGHT uses the full SED to constrain
the stellar populations, an equivalent technique as the one used with
the PEARS dataset.  Comparisons between STARLIGHT ages and stellar
masses and those determined with the chemical enrichment modelling in
\citet{ig09c} are fully consistent within error bars.

\begin{figure}
\includegraphics[width=9cm]{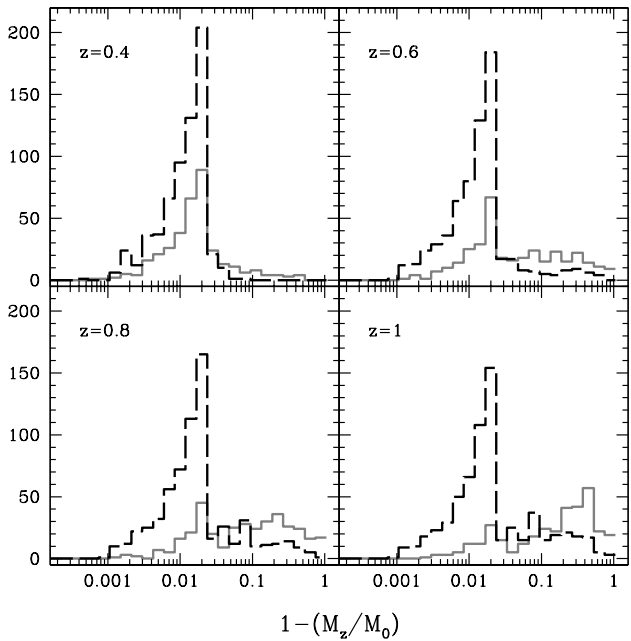}
  
\caption{The predicted fraction of stellar mass growth for our local
  sample is shown when backtracked to four redshift bins, as
  labelled. The histogram only considers galaxies with stellar mass
  M$_s>10^{11}$M$_\odot$. The distribution is shown for the youngest
  (grey, solid line) and oldest (black, dashed line) halves, according
  to the models.}

  \label{fig:DMass}
\end{figure}

\subsection{Backtracking the evolution of local early-type galaxies}

By extracting the star formation histories (SFHs) of our local
galaxies, one can backtrack their evolutionary paths and estimate the
amount of new stellar mass created due to the formation of new stars
as well as the age of the stellar populations at a given redshift. To
minimize systematic effects, we apply the same methodology to both
local and distant galaxies to determine their ages. In
Fig.~\ref{fig:DMass} we show the predicted amount of stellar mass for
the galaxies in our local sample formed since z$\sim$1 according to
their SFHs.

Fig.~\ref{fig:DMass} uses the best fit models from STARLIGHT to
quantify the net increase in stellar mass from recent phases of star
formation. We show the mass growth as the ratio between the stellar
mass already in place at some redshift (M$_z$) and the current mass at
redshift zero (M$_0$) for four redshift bins. The sample is split at
the median in age measured at zero redshift. One can see that the
stellar mass growth at $z\simlt 0.6$ stays well below 10\% for most of
the galaxies, especially for the most massive galaxies, which belong
mostly to the oldest half (black solid lines).

\begin{figure*}
\includegraphics[width=13cm]{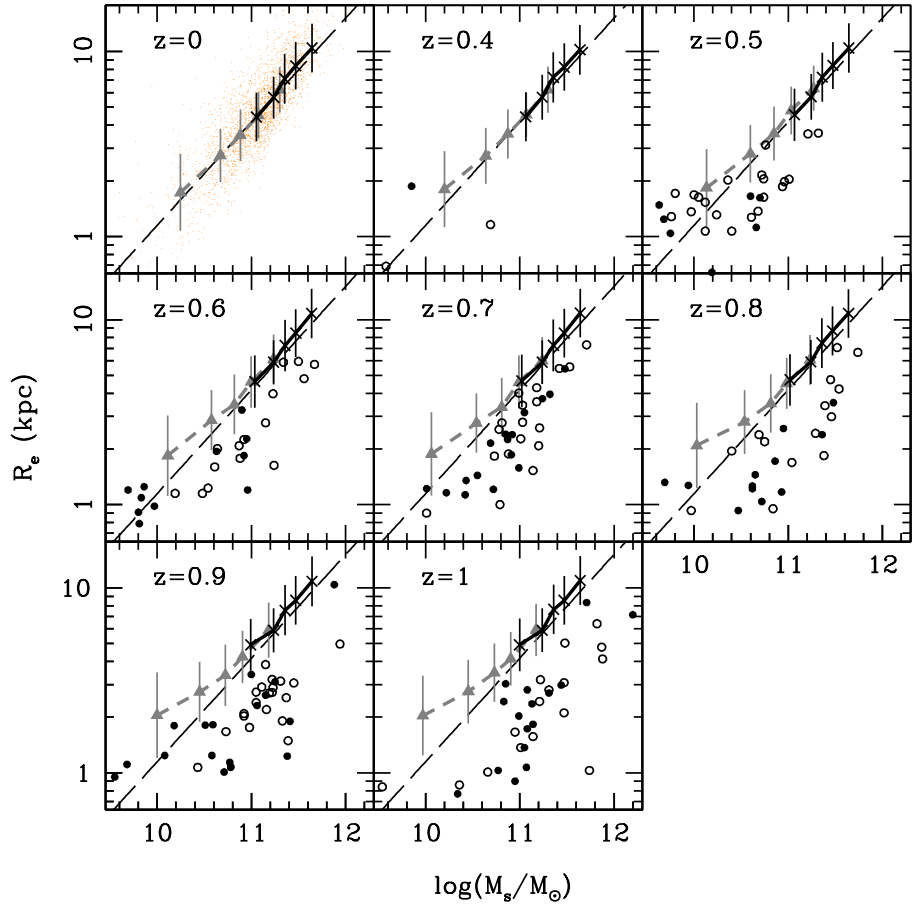}
  
\caption{Comparison of the stellar mass--size relation for the
  extrapolated values of the local sample and the PEARS sample, in
  redshift bins, as labelled. The black crosses (grey triangles)
  represent the oldest (youngest) half in mass-weighted age of the
  local sample extrapolated backwards in time according to the best
  fit star formation history (see text for details). The circles
  correspond to the PEARS sample. Open (solid) circles represent
  galaxies older (younger) than the median value, computed within each
  redshift bin. Note the local sample is corrected regarding the
  evolution of the total stellar mass, but not in size, so that a
  vertical shift should be expected when comparing both samples (see
  figure~\ref{fig:Revol}). At z$=$0 (top-left panel) individual
  galaxies are shown as dots and the local early-type stellar
  mass--size relation from SDSS \citep{she03} is given as a
  long-dashed line in all the panels.}

  \label{fig:Comp}
\end{figure*}

We have applied to all our galaxies in the local relation the
evolution in mass predicted from their SFHs and we have rebuilt the
local stellar mass-size relation taking into account that evolution.
We consider both the change in stellar age (mass-weighted) and the
change in stellar mass of the galaxy.  
 For simplicity, we assume that, within a galaxy, the star
  formation history does not have a radial trend. Our sample does not
  allow us to probe in detail this point, but we note that studies of
  the colour gradient of early-type galaxies at moderate redshift find
  almost always the star formation concentrated in the centre, i.e. in
  a blue core \citep{ig05,ig09a}.
The stellar mass--size relation of our PEARS sample in comparison with
the local sample is shown in Fig.~\ref{fig:Comp}.  The figure shows
how the local stellar mass--size relation will look like at different
redshifts if we correct for the stellar mass evolution. One can see
that the redshifted local stellar mass-size relation changes very
little in the high mass regime. The evolution is more evident at lower
masses, where the galaxies clearly deviate from the local
relationship.

\section{Size evolution}

We are now in a position to explore the size evolution of the
early-type galaxies after accounting for the stellar mass growth due
to new star formation. In fact, the comparison with the {\sl observed}
PEARS sample for similar stellar ages will allow us to determine the
evolution of size at a given stellar mass.

At the top-left corner of Fig.~\ref{fig:Comp}, the local sample is
shown using the same criterion as in Fig.~\ref{fig:Sz0}, with
individual galaxies shown as small dots. We include in that panel the
local trend of SDSS early-type galaxies \citep[long dashed
line,][]{she03}, showing agreement with our local sample,
 except for the most massive end, as discussed in Section 3. 
In the following panels, PEARS individual galaxies appear as solid
(open) circles, with ages younger (older) than the median within each
redshift bin. The standard downsizing trend is apparent in this
figure, with the younger PEARS galaxies having the lowest stellar
masses. If the proposed model in \citet{vdwel08} were correct, with
the youngest galaxies being more extended, at a given stellar mass,
than the older counterparts, one would expect this segregation to be
more evident at higher redshifts, where the effect of lookback time
makes it easier to discriminate with respect to age (i.e. a reduced
age-metallicity degeneracy). However, no clear trend with respect to
galaxy size is found in our data.

\begin{figure}
\includegraphics[width=9cm]{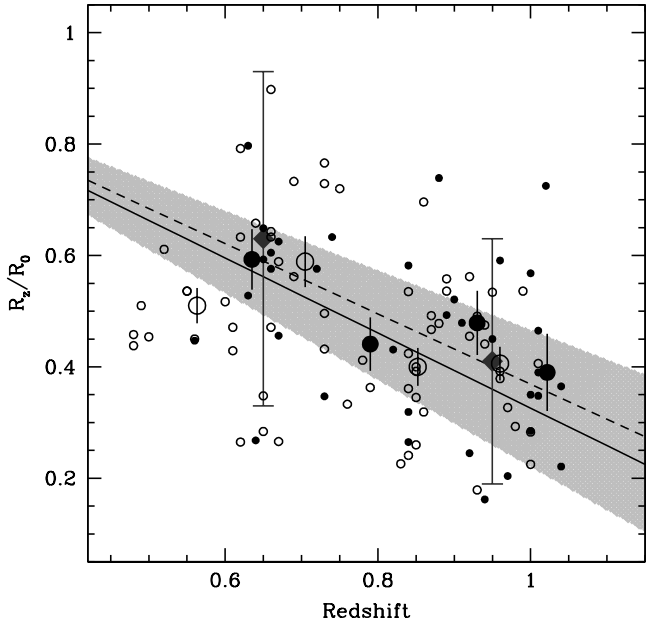}
  
\caption{Size evolution of the PEARS galaxies split according to
  their stellar age. Only galaxies with stellar mass between
  $5\times10^{10}$ and $3\times10^{11}$M$_\odot$ are considered.
  Small open (solid) circles show the evolution for galaxies older
  (younger) than the median within each redshift bin.  The diamonds
  give the size evolution of massive (M$_s\!>\!10^{11}$M$_\odot$)
  spheroids (S\'ersic index n$>\!2.5$) from \citet{truj07}. The error
  bars of the value of \citet{truj07} represent the scatter of their
  sample. The lines represent a linear fit R(z)=R(0)$(1+\gamma z)$ to the
  different galaxy populations: solid line (older subsample) and
  dashed line (younger subsample).The grey solid area is the fit to
  the full population including the 68\% confidence level.}

  \label{fig:Revol}
\end{figure}

Our best fit models for the local sample predict very small stellar
mass changes (see Fig.~\ref{fig:DMass}), at levels that correspond to
$\Delta\log$M$_s\simlt$0.05~dex along the horizontal direction in
Fig.~\ref{fig:Comp}. The comparison with the PEARS sample shows that
there is a noticeable ``vertical'' evolution (i.e. change in
size). This one can be illustrated by comparing the (redshift zero)
size of the local galaxies with the observed size of the PEARS
galaxies, within subsamples of the same stellar
age. Fig.~\ref{fig:Revol} shows the size evolution for galaxies with
stellar mass in the range
$5\!\times\!10^{10}\!<$M$_s$/M$_\odot\!<3\!\times\!10^{11}$. We have
fitted our evolution using the following parametrization:
R(z)=R(0)$(1+\gamma z)$ with R(0) the size obtained from the local stellar
mass-size relation \citep[]{she03}. Our data is compatible with the
following value $\gamma$=-0.657$\pm$0.122 for the full galaxy sample
and with $\gamma$=-0.631$\pm$0.176 for the young subsample and
$\gamma$=-0.674$\pm$0.160 for the older subsample (uncertainties
quoted at the 68\% confidence level).

Fig.~\ref{fig:Revol} shows that the size evolution is significant, in
agreement with, e.g.  \citet{truj07}, with galaxies at $z\sim 1$ being
$\sim$50\% smaller in size than their local counterparts. Notice the
little difference between the trend of the sample segregated with
respect to age (large open/solid grey circles). This is one of the
most important result of this work and implies that the amount of size
evolution that elliptical galaxies suffers since z$\sim$1 is
independent of the age of the galaxies at each redshift interval. This
means that, the full population of elliptical galaxies, independently
of its level of star formation, experiences a similar evolutionary
mechanism for assembly. This is once more a result in contradiction
with the idea that younger galaxies at all redshifts are born with
significantly larger sizes than their older massive counterparts. In
other words, our results point out to a similar displacement in the
stellar mass--size relation of all the galaxies in the sample
(independently of their age).

\section{Constraining the different evolutionary paths of the
  elliptical galaxies since z$\sim$1}

In this section we explore the current most likely scenarios proposed 
to explain the evolution of elliptical galaxies on the mass--size
plane. We use the results obtained here and in previous papers to
constrain those scenarios.

In what follows we consider that both the size and the stellar mass
growth of the elliptical galaxies can be described as the contribution
of three different processes: i) formation of new stars in the
galaxies as a result of gas consumption, ii) accretion of already
formed stars from merging of different subunits and iii) gas ejection
from the activity of either an AGN and/or supernova galactic winds. We
parameterize the effect of these three processes in the mass and size of the
galaxies as follows:

\begin{equation}
\Delta M_s= \Delta M_{s, SF}+ \Delta M_{s,acc}
\end{equation}
\begin{equation}
\Delta r_{e}= \Delta r_{e, SF}+ \Delta r_{e,acc} + \Delta r_{e,agn}
\end{equation}

\noindent
with $\Delta M_{s, SF}$ and $\Delta M_{s,acc}$ representing the
increase of the stellar mass due to star formation and by accretion of
new stars into the galaxies, respectively. $\Delta r_{e, SF}$, $\Delta
r_{e,acc}$ and $\Delta r_{e,agn}$ correspond to the increase in size
by star formation, accreted stars and by expansion due to galactic
winds either created by the effect of a central AGN or supernovae
explosions.

\subsection{Observational facts}

The results of this paper show that $\Delta M_{s, SF}$ is very small
(i.e. $\Delta M_{s, SF} << M_{s}$) and also that the evolution of the
size of the galaxies is quite independent of the age of their stellar
population, so $\Delta r_{e}({\rm old})\sim\Delta r_{e}({\rm young})$. 
Due to the little increase in the stellar mass due to in-situ star
formation we assume from now on, to simplify the discussion, that, if
any, $\Delta M_s \approx \Delta M_{s,acc}$ for the elliptical galaxies
since z$\sim$1.

\subsection{Puffing up model: AGN and/or supernova galactic winds effects}

\citet{fan08,fan10} have proposed a mechanism based on the removal of
gas as result of AGN activity to explain the size growth of early-type
galaxies. According to these authors, the rapid expulsion of large
amounts of gas by quasar winds destabilizes the galaxy structure in
the inner, baryon-dominated regions, and leads to a more expanded
stellar distribution. A similar idea -- but based on the gas expulsion
associated to stellar evolution -- has been proposed by
\citet[]{dam09}. The prediction from the puffing up model can be
parameterized as follows:

\begin{eqnarray}
\Delta M_{s, SF}=0 \\
\Delta M_{s,acc}=0 \\
\Delta M_s=0 \\
\Delta r_{e, SF}=0 \\
\Delta r_{e,acc}=0 \\
\Delta r_{e}=\Delta r_{e,agn}
\end{eqnarray}

In other words, all the galaxies in the stellar mass-size relation
should just evolve vertically in this relation without any increase of
stellar mass. Consequently, the size evolution we observe at fixed
stellar mass should be directly interpreted as the total size
evolution of the galaxies.

This model agrees with observations at predicting a small formation of
new stars due to the removal of gas from the galaxies. In addition,
this model fits well with the lack of evidence of significant
evolution in the number density of massive ellipticals since
z$\sim$1. However, we find that our data is in conflict with the model
in several aspects. First, according to the \citet{fan08} model, after
the formation of the compact structure, the AGN activity will remove
the gas, triggering a fast growth process ($\sim$20-30 Myr based on
recent simulations\footnote{In the case of supernova winds an
  important mass loss event could last even $\sim$0.5-1 Gyr.};
\citet[]{rag11}. This would imply that galaxies with stellar
populations older than $\sim$1 Gyr should be already located in the
local stellar mass-size relation.  This is not what our data shows. We
have galaxies (old and young) at the same distance from the local
relation at all redshifts. For instance, at z=1 the mass-weighted age
of our sample is 3.9~Gyr for the old subsample and 3.5~Gyr for the
young subsample. We can consequently assure that the mechanism that is
operating in the size evolution of our galaxies does not know about
the age of the stellar populations.  This is in contradiction with the
puffing up model. In addition, a natural prediction from the puffing
up model is that the scatter of the stellar mass-size relation will
increase with redshift \citep{fan10}, with some galaxies already in
place on the local relation and others still in a very compact
phase. We do not observe any increase in the scatter of the stellar
mass-size relation with redshift in our data.

\subsection{Major dry mergers}

Major mergers (i.e. mergers of galaxies with similar mass) were first
considered as one the likely paths for size growth in elliptical
galaxies. Major dry mergers can increase the size in a way almost
directly proportional to the mass increase
\citep[e.g.][]{cio01,nip03,boy06,naab07}. This evolution is not strong
enough to be compatible with the low number of major mergers observed
at least since z$\sim$1 \citep[]{bun09,wid09,der09,lop10},  as well as
with recent numerical simulations \citep{ks09}. For this
reason, we will not consider this mechanism further.

\subsection{Minor dry merging}

Another possible scenario for elliptical galaxy growth involves to
minor mergers on parabolic orbits
\citep[e.g.][]{kb06,mal06,naab09,hop09b}. Through this channel, the new
accreted stars as well as the redistribution of stars in the main
galaxy, preferentially populate the outer region of the objects.  For
this reason, this mechanism has been considered a very efficient way
of size growth. \citet[]{fan10}, following \citet[]{naab09}, show
that the fractional variation of the gravitational radius and the
velocity dispersion of the main galaxy before (i) and after (f) a
minor merger is:

\begin{eqnarray}
\frac{R_{f}}{R_{i}}=\frac{(1+\eta)^2}{1+\eta^{2-\alpha}} \\
\frac{\sigma_f^2}{\sigma_i^2}=\frac{1+\eta^{2-\alpha}}{1+\eta},
\end{eqnarray}

\noindent
with $\eta$ defined as M$_f$=M$_i$(1+$\eta$) and $\alpha$ representing
the exponent of the local stellar mass-size relation
(R=bM$_\star^\alpha$). \citet{she03} proposes $\alpha\approx$0.56
with b=2.88$\times$10$^{-6+11\alpha}$ (in units of
10$^{11}$M$_{\sun}$). On what follows, we implicitly assume that the
gravitational radius is proportional to the effective radius of our
galaxies. This is only strictly correct as long as the galaxies do
not change the shape of their surface brightness profiles during the
minor merger process.

It can be shown that after N mergers of equal mass ratio $\eta$, the
final mass, velocity dispersion and radius can be written as:

\begin{eqnarray}
\frac{R_{f}}{R_{i}}=\Big[\frac{(1+\eta)^2}{1+\eta^{2-\alpha}}\Big]^N\\
\frac{M_{f}}{M_{i}}=(1+\eta)^N\\
\frac{\sigma_f^2}{\sigma_i^2}=\Big[\frac{1+\eta^{2-\alpha}}{1+\eta}\Big]^N
\end{eqnarray}

\noindent
We can now make an estimation of the number, N, of minor mergers a
galaxy requires in order to reach the present stellar mass-size
relation. The final size of the galaxy can be written in terms of the
initial size, the size evolution at a fixed stellar mass (provided by
the observations) and the difference in stellar mass as follows:

\begin{equation}
\log R_{f}\equiv \log R_{i}+\Delta\log R\Big|_{M_{i,fixed}}+\alpha\log(M_f/M_i).
\end{equation}

\noindent
The evolution at fixed stellar mass for different redshifts is
determined by the size evolution found in our data 
$\Delta\log R|_{M_{i,fixed}}=-\log(1+\gamma z)$, with
$\gamma=-0.657\pm 0.122$. Using Eq.~(13), (14) and (16) we find
for the number of minor mergers:

\begin{equation}
N = \frac{-\log (1+\gamma z)}{\log\Big[\frac{(1+\eta)^{2-\alpha}}{1+\eta^{2-\alpha}}\Big]}.
\end{equation}

\begin{figure}
\includegraphics[width=9cm]{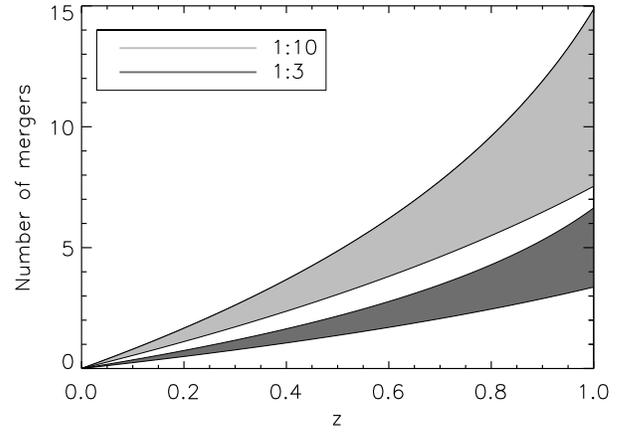}
\caption{Predicted number of minor mergers as a function of redshift
  according to the observed size evolution in our data. The grey
  regions represent the number of mergers within a 1$\sigma$
  uncertainty, for two choices of the mass ratio ($\eta$). As
  expected, the number of minor mergers is a function of $\eta$, being
  larger for smaller ratios.}
   \label{fig:nummergers}
\end{figure}

\noindent
We show in Fig.~\ref{fig:nummergers} this number as a function of
redshift for two different values of $\eta$: 1/3 and 1/10. As
expected, the number of minor mergers is a function both of 
redshift and the mass increase per merger, $\eta$. We can use these estimations in
the number of minor mergers as a function of redshift to determine the
increase in size, stellar mass and velocity dispersion that individual
galaxies suffer if their evolution is dictated by the minor dry
merging scenario. This is quantified in
Table~\ref{tab:minormerging}. Since z$\sim$0.8, individual objects
undergo size growth by a factor around 3.5, whereas the stellar mass
grows a factor around 2.5. As expected, the velocity dispersion of the
individual galaxies decreases with time due to the minor merging. The
evolution is, however, very mild. We can compare this evolution with
the observed values found in \citet{ct09}. The comparison is not
straightforward as those authors measure the velocity dispersion
increase at a fixed stellar mass, whereas we have followed the
evolution for {\sl individual} galaxies and we observe that the
increase in stellar mass is not negligible. \citet{ct09} find a
velocity dispersion change ranging from 0.84 to 0.90 since z=0.8 which
is in good correspondence to the observed values predicted here from
the observed size evolution.

\begin{table*}

\caption{Number of minor mergers, size, mass and velocity dispersion evolution for individual galaxies
according to the minor dry merging model compatible with our data.}

\label{tab:minormerging}
\begin{center}
\begin{tabular}{ccccc}
\hline
Redshift & $<$N$>$($\pm$1$\sigma$) &  R$_{f}$/R$_{i}$($\pm$1$\sigma$)  & M$_{f}$/M$_{i}$($\pm$1$\sigma$)
& $\sigma$$_{f}$/$\sigma$$_{i}$($\pm$1$\sigma$)\\
\hline
& & $\eta$=1/3 & & \\
\hline
     0.2 &    0.62$\pm$0.12 &	 1.27$\pm$0.06 &    1.20$\pm$0.04  &  0.97$\pm$0.01 \\
     0.4 &    1.35$\pm$0.29 &	 1.70$\pm$0.19 &    1.48$\pm$0.12  &  0.93$\pm$0.01 \\
     0.6 &    2.24$\pm$0.53 &	 2.44$\pm$0.50 &    1.93$\pm$0.29  &  0.89$\pm$0.02 \\
     0.8 &    3.38$\pm$0.92 &	 3.95$\pm$1.35 &    2.73$\pm$0.71  &  0.84$\pm$0.04 \\
     1.0 &    5.00$\pm$1.64 &	 8.44$\pm$4.74 &    4.70$\pm$2.06  &  0.78$\pm$0.06 \\
\hline
& & $\eta$=1/10 & & \\
\hline
     0.2 &   1.40$\pm$0.28 &   1.24$\pm$0.05 & 1.14$\pm$0.03  &  0.96$\pm$0.01	\\
     0.4 &   3.02$\pm$0.65 &   1.61$\pm$0.16 & 1.33$\pm$0.08  &  0.91$\pm$0.02	\\
     0.6 &   5.00$\pm$1.20 &   2.21$\pm$0.40 & 1.62$\pm$0.18  &  0.86$\pm$0.03	\\
     0.8 &   7.56$\pm$2.05 &   3.39$\pm$1.04 & 2.09$\pm$0.40  &  0.80$\pm$0.05	\\
     1.0 &  11.20$\pm$3.66 &   6.61$\pm$3.39 & 3.09$\pm$1.04  &  0.72$\pm$0.08	\\
\hline
\end{tabular}
\end{center}
\end{table*}

\section{Discussion and Conclusions}

In this paper we have explored how the local stellar mass-size
relation of elliptical galaxies has been built up since z$\sim$1.  We
have compiled a sample of visually classified elliptical galaxies
since z$\sim$1 from the GOODS datasets, as well as from SDSS data. All
our galaxies have spectroscopic data that enable a robust constraint
of the age and mass of the underlying stellar populations. Both the
study of the fossil record in the local relation as well as the
analysis of the stellar mass-size relation evolution with redshift
agree on an evolutionary mechanism that is mostly insensitive to
the age of the stellar populations of the galaxies at all redshifts.

We do not find any clear evidence for a progressive buildup of the
local stellar mass--size relation following a bottom-up sequence. In
other words, we do not observe that the smaller galaxies, at fixed
stellar mass, are generally older than the larger galaxies. On the
contrary, the local stellar mass-size relation seems to be in place
(with a similar slope and scatter, at least since z$\sim$1) but with
all the galaxies presenting a "vertical drift" towards smaller sizes.

The analysis of our data rejects the puffing up scenario which
proposes that the growth in size is due to the rapid expulsion of
large amounts of gas by the effect of an AGN or supernovae-driven
winds.  In fact, two key predictions of this model, the increase of
the scatter in size with redshift at fixed mass, and the absence of old
galaxies with small sizes are not observed in our data. Our data,
however, is not in conflict with an increase of the galaxy sizes
through minor merging. 
 Minor merging has been also favoured by studies using different
  methods and with samples at higher redshift
  \citep[i.e.][]{bez09,vdk10}.
Under this hypothesis we have calculated the number of minor mergers
that would be necessary to build the local stellar mass-size relation
in agreement with the observed size evolution. Since z=0.8, we find
$\sim$3$\pm$1 mergers with ratio 1:3 or $\sim$8$\pm$2 with ratio 1:10.

The data analyzed in this work together with the evidence collected in
recent papers \citep[e.g.][]{kav09,shan10,nier11} only leaves the minor
merging scenario as a viable mechanism for the size increase of
elliptical galaxies at least since z$\sim$1. Proving ultimately,
however, that elliptical galaxies grow by minor merging will require a
direct quantification on the minor merger events found in high
redshift galaxies and an exploration of the age and metallicity
gradients of the stellar population in local elliptical galaxies.

\section*{Acknowledgments}

 We would like to thank S. Khochfar, C. Tortora, F. Shankar and
S. Kaviraj for useful comments. The anonymous referee is also
acknowledged for his/her constructive feedback.
IT is a Ram\'on y Cajal Fellow of the Spanish Ministry of Science and
Innovation. IF acknowledges a grant from the Royal Society and support
from the IAC to carry out this research project.  IGR acknowledges a
grant from the Spanish Secretar\'\i a General de Universidades of the
Ministry of Education, in the frame of its programme to promote the
mobility of Spanish researchers to foreign centers.  This work has
been supported by the ``Programa Nacional de Astronom\'\i a y
Astrof\'\i sica'' of the Spanish Ministry of Science and Innovation
under grant AYA2010-21322-C03-02.



\begin{thebibliography}{}
\bibitem[\protect\citeauthoryear{Abazajian et al.}{2009}]{dr7} 
Abazajian, K.~N., et al. 2009, ApJS, 182, 543

\bibitem[\protect\citeauthoryear{Adelman-McCarthy et al.}{2006}]{sdssdr4} 
Adelman-McCarthy, J.~K., et al. 2006, ApJSS, 162, 38  

\bibitem[\protect\citeauthoryear{Adelman-McCarthy et al.}{2008}]{sdssdr6} 
Adelman-McCarthy, J.~K., et al. 2008, ApJSS, 175, 297

\bibitem[\protect\citeauthoryear{Banerji et al.}{2010}]{ban10} 
Banerji, M., Ferreras, I., Abdalla, F.~B., Hewett, P., Lahav, O., 
MNRAS, 402, 2264

\bibitem[\protect\citeauthoryear{Bernardi et al.}{2005}]{ber05} 
Bernardi, M., Sheth, R.~K., Nichol, R.~C., Schneider, D.~P., 
Brinkmann, J., 2005, AJ, 129, 61 

\bibitem[\protect\citeauthoryear{Bertin et al.}{2002}]{bert02} 
Bertin, G., Ciotti, L., Del Principe, M., 2002, A\&A, 386, 149

\bibitem[\protect\citeauthoryear{Bezanson et al.}{2009}]{bez09} 
Bezanson, R., et al., 2009, ApJ, 697, 1290

\bibitem[\protect\citeauthoryear{Birnboim \& Dekel}{2003}]{bd03}
Birnboim, Y., Dekel, A., 2003, MNRAS, 345, 349

\bibitem[\protect\citeauthoryear{Boylan-Kolchin et al.}{2006}]{boy06} 
Boylan-Kolchin, M., Ma. C.-P., Quataert, E., 2006, MNRAS, 369, 1081

\bibitem[\protect\citeauthoryear{Blanton et al.}{2005}]{blan05}
Blanton, M. R., Eisenstein, D., Hogg, D. W., Schlegel, D. J., Brinkmann, J.,
2005, ApJ, 629, 143

\bibitem[\protect\citeauthoryear{Bruzual \& Charlot}{2003}]{BC03} 
Bruzual, G., Charlot, S., 2003, MNRAS, 344, 1000

\bibitem[\protect\citeauthoryear{Bruzual}{2007}]{BC07} 
Bruzual, G. 2007, IAU No. 241 Symp. Procs. "Stellar populations as
  building blocks of galaxies", eds. A. Vazdekis and R.~F. Peletier,
  Cambridge, arXiv:astro-ph/0703052

\bibitem[\protect\citeauthoryear{Buitrago et al.}{2008}]{bui08}
Buitrago, F., Trujillo, I., Conselice, C.~J., Bouwens, R.~J., 
Dickinson, M., Yan, H., 2008, ApJ, 687, L61

\bibitem[\protect\citeauthoryear{Bundy et al.}{2009}]{bun09}
Bundy, K., Fukugita, M., Ellis, R.S. et al., 2009, ApJ, 697, 1369

\bibitem[\protect\citeauthoryear{Carrasco et al.}{2010}]{car10} 
Carrasco, E.R., Conselice, C.J., Trujillo, I., 2010, MNRAS, 405, 2253

\bibitem[\protect\citeauthoryear{Cappellari et al.}{2006}]{cap06} 
Cappellari, M., et al. 2006, MNRAS

\bibitem[\protect\citeauthoryear{Cardelli et al.}{1989}]{CCM89} 
Cardelli, J.~A., Clayton, G.~C., Mathis, J.~S., 1989, ApJ, 345, 245 

\bibitem[\protect\citeauthoryear{Cenarro \& Trujillo}{2009}]{ct09}
Cenarro, J., Trujillo, I., 2009, ApJ, 696, L43

\bibitem[\protect\citeauthoryear{Cimatti et al.}{2008}]{cim08}
Cimatti, A., et al. 2008, A\& A, 482, 21

\bibitem[\protect\citeauthoryear{Conselice et al.}{2007}]{cc07}
Conselice, C.~J. et al. 2007, MNRAS, 381, 962

\bibitem[\protect\citeauthoryear{Cid Fernandes et al.}{2005}]{CF05} 
Cid Fernandes, R., Mateus, A., Sodr\'e, L., Stasinska, G., 
Gomes, J.~M., 2005, MNRAS, 356, 270

\bibitem[\protect\citeauthoryear{Ciotti \& van Albada }{2001}]{cio01}
Ciotti, L., van Albada, T. S., 2001, ApJ, 552, L13 

\bibitem[\protect\citeauthoryear{Daddi et al.}{2005}]{dad05} 
Daddi, E., et al. 2005, ApJ, 626, 680

\bibitem[\protect\citeauthoryear{Damjanov et al.}{2009}]{dam09} 
Damjanov, I., et al. 2009, ApJ, 695, 101

\bibitem[\protect\citeauthoryear{de Ravel et al.}{2009}]{der09} 
de Ravel, L., Le F\`evre, O., Tresse, L., et al. 2009, A\&A, 498, 379

\bibitem[\protect\citeauthoryear{Dekel et al.}{2009}]{dek09} 
Dekel, et al., 2009, Nature 457, 451 

\bibitem[\protect\citeauthoryear{Fan et al.}{2008}]{fan08}
Fan, L., Lapi, A., De Zotti, G., Danese, L. 2008, ApJ, 689, L101

\bibitem[\protect\citeauthoryear{Fan et al.}{2010}]{fan10}
Fan, L., Lapi, A., Bressan, A., Bernardi, M., De Zotti, G.,
Danese, L. 2010, ApJ, 718, 1460

\bibitem[\protect\citeauthoryear{Ferreras et al.}{2005}]{ig05}
Ferreras, I., Lisker, T., Carollo, C.~M., Lilly, S.~J.,
Mobasher, B., 2005, ApJ, 635, 243

\bibitem[\protect\citeauthoryear{Ferreras et al.}{2005}]{fsw05}
Ferreras, I., Saha, P., Williams, L.~L.~R. 2005, ApJ, 623, L5

\bibitem[\protect\citeauthoryear{Ferreras et al.}{2009a}]{ig09a}
Ferreras, I., Lisker, T., Pasquali, A., Kaviraj, S. 2009a, MNRAS, 395, 554

\bibitem[\protect\citeauthoryear{Ferreras et al.}{2009b}]{ig09b}
Ferreras, I., Lisker, T., Pasquali, A., Khochfar, S., 
Kaviraj, S. 2009b, MNRAS, 396, 1573

\bibitem[\protect\citeauthoryear{Ferreras et al.}{2009c}]{ig09c}
Ferreras, I. et al. 2009c, ApJ, 706, 158

\bibitem[\protect\citeauthoryear{Fontana et al.}{2006}]{fon06}
Fontana, A., et al. 2006, A\&A, 459, 745

\bibitem[\protect\citeauthoryear{Gallazzi et al.}{2005}]{gal05}
Gallazzi, A., Charlot, S., Brinchmann, J., White, S.~D.~M.,
Tremonti, C.~A., 2005, MNRAS, 362, 41

\bibitem[\protect\citeauthoryear{Giavalisco et al.}{2004}]{goods}
Giavalisco, M., et al. 2004, ApJ, 600, L93

\bibitem[\protect\citeauthoryear{Graves et al.}{2008}]{grav08}
Graves, G.~F., Faber, S.~M., Schiavon, R.~P., 2009, ApJ, 693, 486

\bibitem[\protect\citeauthoryear{Guo et al.}{2009}]{guo09} 
Guo, Y., et al. 2009, MNRAS, 398, 1129

\bibitem[\protect\citeauthoryear{Hopkins et al.}{2009}]{hop09}
Hopkins, P.~F., Bundy, K., Murray, N., Quataert, E., Lauer, T.~R.,
Ma. C., 2009, MNRAS, 398, 898

\bibitem[\protect\citeauthoryear{Hopkins et al.}{2009b}]{hop09b}
Hopkins, P.~F., Hernquist, L., Cox, T.~J., Keres, D., Wuyts, S., 
2009b, ApJ, 691, 1424

\bibitem[\protect\citeauthoryear{J\o rgensen et al.}{1995}]{Jo95} 
J\o rgensen, I., Franx, M.,  Kj\ae rgaard, P. 1995, MNRAS, 276, 1341

\bibitem[\protect\citeauthoryear{Kaviraj et al.}{2009}]{kav09} 
Kaviraj, S., et al. 2009, MNRAS, 394, 1713

\bibitem[\protect\citeauthoryear{Kere\u s et al.}{2009}]{ker09} 
Kere\u s, D., Katz, N., Fardal, M., Dav\'e, R., Weinberg, D.~H.
2009, MNRAS, 395, 160

\bibitem[\protect\citeauthoryear{Khochfar \& Burkert}{2006}]{kb06}  
Khochfar, S., Burkert, A., 2006, A\& A, 445, 403

\bibitem[\protect\citeauthoryear{Khochfar \& Silk}{2006}]{ks06}  
Khochfar, S., Silk, J., 2006, MNRAS, 370, 902

\bibitem[\protect\citeauthoryear{Khochfar \& Silk}{2009}]{ks09} 
Khochfar, S., Silk, J., 2009, MNRAS, 397, 506

\bibitem[\protect\citeauthoryear{Kroupa}{2001}]{K01} 
Kroupa, P. 2001, MNRAS, 322, 231

\bibitem[\protect\citeauthoryear{Leier et al.}{2011}]{leier11} 
Leier, D., Ferreras, I., Saha, P., Falco, E., 2011, ApJ, arXiv:1102.3433

\bibitem[\protect\citeauthoryear{Lintott et al.}{2011}]{lint11} 
Lintott, C., et al., 2011, MNRAS, 410, 166

\bibitem[\protect\citeauthoryear{Longhetti et al.}{2007}]{lon07} 
Longhetti, M. et al. 2007, MNRAS, 374, 614

\bibitem[\protect\citeauthoryear{L\'opez-Sanjuan et al.}{2010}]{lop10} 
L\'opez-Sanjuan, C., Balcells, M., P\'erez-Gonz\'alez, P. G., et al. 2010, ApJ, 710, 1170

\bibitem[\protect\citeauthoryear{La Barbera et al.}{2010}]{LB10} 
La Barbera, F., de Carvalho, R.~R., de la Rosa, I.~G., Lopes, P.~A.~A., 
Kohl-Moreira, J.~L., Capelato, H.~V., 2010, MNRAS, 408, 1313

\bibitem[\protect\citeauthoryear{Maller et al.}{2006}]{mal06}
Maller, A. H., et al. 2006, ApJ, 647, 763

\bibitem[\protect\citeauthoryear{McIntosh et al.}{2005}]{mci05}
McIntosh D. et al., 2005, ApJ, 632, 191

\bibitem[\protect\citeauthoryear{Moster et al.}{2010}]{mos10} 
Moster, B.~P., Somerville, R.~S., Maulbetsch, C., van den Bosch, F.~C., 
Macci\`o, A.~V., Naab, T., Oser, L, 2010, ApJ, 710, 903

\bibitem[\protect\citeauthoryear{Naab et al.}{2007}]{naab07}
Naab, T., Johansson, P.~H.,Ostriker, J.~P., Efstathiou, G., 2007, ApJ, 658, 710

\bibitem[\protect\citeauthoryear{Naab et al.}{2009}]{naab09}
Naab, T., Johansson, P.~H., Ostriker, J.~P., 2009, ApJ, 699, L178

\bibitem[\protect\citeauthoryear{Nair \& Abraham}{2010}]{na10}
Nair, P.~B., Abraham, R.~G., 2010, ApJSS, 186, 427 

\bibitem[\protect\citeauthoryear{Napolitano et al.}{2010}]{nap10}
Napolitano, N.~R., Romanowsky, A.~J., Tortora, C., 2010, MNRAS, 405, 2351

\bibitem[\protect\citeauthoryear{Nierenberg et al.}{2011}]{nier11}
Nierenberg, A.~M., et al. 2011, ApJ, 731, 44

\bibitem[\protect\citeauthoryear{Nipoti et al.}{2010}]{nip03}
Nipoti, C., Londrillo, P., Ciotti, L., MNRAS, 342, 501

\bibitem[\protect\citeauthoryear{Ragone-Figueroa \& Granato}{2011}]{rag11}
Ragone-Figueroa, C., Granato, G. L., 2011, arXiv:1101.4947

\bibitem[\protect\citeauthoryear{Rogers et al.}{2010}]{rog10}
Rogers, B., Ferreras, I., Pasquali, A., Bernardi, M., Lahav, O.,
Kaviraj, S. 2010, MNRAS, 405, 329

\bibitem[\protect\citeauthoryear{S\'anchez-Bl\'azquez et al.}{2006}]{SB06} 
S\'anchez-Bl\'azquez, P., Peletier, R.~F., Jim\'enez-Vicente, J., 
Cardiel, N., Cenarro, A.~J., Falc\'on-Barroso, J., Gorgas, J., 
Selam, S., Vazdekis, A. 2006, MNRAS, 371, 703

\bibitem[\protect\citeauthoryear{Saracco et al.}{2011}]{sar11}
Saracco, P., Longhetti, M., Gargiulo, A., 2011, MNRAS, in press

\bibitem[\protect\citeauthoryear{Schlegel et al.}{1998}]{Sch98} 
Schlegel, D.J., Finkheiner, D.P., Davis, M. 1998, ApJ, 500, 525
  
\bibitem[\protect\citeauthoryear{Shankar \& Bernardi}{2009}]{shan09}
Shankar, F., Bernardi, M., 2009, MNRAS, 396, L76

\bibitem[\protect\citeauthoryear{Shankar et al.}{2010}]{shan10}
Shankar, F., et al., 2010, MNRAS, 403, 117

\bibitem[\protect\citeauthoryear{Shen et al.}{2003}]{she03}
Shen, S., Mo, H.J., White, S. D. M., Blanton, M. R., Kauffmann, G., Voges, W., 
Brinkmann, J., Csabai, I., 2003, MNRAS, 343, 978

\bibitem[\protect\citeauthoryear{Taylor et al.}{2010}]{tay10}
Taylor, E.~N., Franx, M., Glazebrook, K., Brinchmann, J., 
van der Wel, A., van Dokkum, P.~G., 2009, ApJ, arXiv:0907.4766

\bibitem[\protect\citeauthoryear{Tortora et al.}{2009}]{tor09}
Tortora, C., Napolitano, N.~R., Romanowsky, A.~J., Capaccioli, M.,
Covone, G., 2009, MNRAS, 396, 1132

\bibitem[\protect\citeauthoryear{Treu et al.}{2005}]{treu05}
Treu, T., et al. 2005, ApJ, 633, 174
 
\bibitem[\protect\citeauthoryear{Trujillo et al.}{2004}]{truj04}
Trujillo I. et al., 2004, ApJ, 604, 521

\bibitem[\protect\citeauthoryear{Trujillo et al.}{2006a}]{truj06a}
Trujillo, I., et al. 2006, ApJ, 650, 18

\bibitem[\protect\citeauthoryear{Trujillo et al.}{2006b}]{truj06b}
Trujillo, I. et al. 2006, MNRAS, 373, L36

\bibitem[\protect\citeauthoryear{Trujillo et al.}{2007}]{truj07}
Trujillo, I., Conselice, C.~J., Bundy, K., Cooper, M.~C., Eisenhardt, P., 
Ellis, R.~S., 2007, MNRAS, 382, 109

\bibitem[\protect\citeauthoryear{Trujillo et al.}{2009}]{truj09}
Trujillo, I., Cenarro, A.~J., de Lorenzo-C\'aceres, A., Vazdekis, A., 
de la Rosa, I.~G., Cava, A. 2009, ApJ, 692, L118

\bibitem[\protect\citeauthoryear{Valentinuzzi et al.}{2010}]{val10}
Valentinuzzi, T. et al. 2010, ApJ, 712, 226

\bibitem[\protect\citeauthoryear{van der Wel et al.}{2005}]{vdwel05}
van der Wel, A., Franx, M., van Dokkum, P.~G., Rix, H.-W., 
Illingworth, G.~D., Rosati, P., 2005, ApJ, 631, 145

\bibitem[\protect\citeauthoryear{van der Wel et al.}{2008}]{vdwel08}
van der Wel, A., Holden, B.~P., Zirm, A.~W., Franx, M., Rettura, A.
Illingworth, G.~D., Ford, H.~C., 2008, ApJ, 688, 48

\bibitem[\protect\citeauthoryear{van der Wel et al.}{2009}]{vdwel09}
van der Wel, A., Bell, E.~F., van den Bosch, F.~C., Gallazzi, A., 
Rix, H.-W., 2009, ApJ, 698, 1232

\bibitem[\protect\citeauthoryear{van Dokkum et al.}{2008}]{vdk08}
van Dokkum, P.~G., et al. 2008, ApJ, 677, L5

\bibitem[\protect\citeauthoryear{van Dokkum et al.}{2010}]{vdk10}
van Dokkum, P.~G., et al. 2010, ApJ, 709, 1018

\bibitem[\protect\citeauthoryear{Vazdekis et al.}{2010}]{Va10} 
Vazdekis, A.; S\'anchez-Bl\'azquez, P.; Falc\'on-Barroso, J.; 
Cenarro, J.; Beasley, M.~A.; Cardiel, N.; Gorgas, J.; 
Peletier, R.~F. 2010, MNRAS, 404, 1639

\bibitem[\protect\citeauthoryear{Wild et al.}{2009}]{wid09} 
Wild, V., Walcher, C. J., Johanson, P. H., et al. 2009, MNRAS, 395, 144

\bibitem[\protect\citeauthoryear{York et al.}{2000}]{sdss} 
York, D.~G. et al. 2000, AJ, 120, 1579

\bibitem[\protect\citeauthoryear{Zirm et al.}{2007}]{zirm07} 
Zirm, A.~W. et al. 2007, ApJ, 656, 66
  
\end{thebibliography}
\end{document}